\documentclass[aps,pre,twocolumn,showpacs,floatfix]{revtex4}

\usepackage{epsfig}
\usepackage{amsmath}
\usepackage{amssymb}
\usepackage{multirow}
\usepackage{rotating} 
\usepackage{color}
\usepackage{graphicx}

\newcommand{\eg}[0]{\textit{e.g.}}
\newcommand{\ie}[0]{\textit{i.e.}}
\newcommand{\ignore}[1]{}

\newcommand{\ELneighX}[0]{M}
\newcommand{\ELneigh}[1]{\ELneighX(#1)}
\newcommand{\kcalmol}[0]{\frac{\text{kcal}}{\text{mol}}}

\begin{document}

\title{Efficient exploration of discrete energy landscapes}
\author{Martin Mann} 
\affiliation{Bioinformatics Group, University of Freiburg,
	Georges-K\"ohler-Allee 106, D-79110 Freiburg, Germany}

\author{Konstantin Klemm}
\affiliation{Bioinformatics Group, Institute for Computer Science,
University of Leipzig, H\"artelstra\ss{}e 16-18, D-04107 Leipzig, Germany
}
	
\date{\today} 

\begin{abstract}
Many physical and chemical processes, such as folding of biopolymers, are best
described as dynamics on large combinatorial energy landscapes. A concise
approximate description of the dynamics is obtained by partitioning the
micro-states of the landscape into macro-states. Since most landscapes of
interest are not tractable analytically, the probabilities of transitions
between macro-states need to be extracted numerically from the microscopic ones,
typically by full enumeration of the state space or approximations using the
Arrhenius law. Here we propose to approximate transition probabilities by a
Markov chain Monte-Carlo method. For landscapes of the number partitioning
problem and an RNA switch molecule we show that the method allows for accurate
probability estimates with significantly reduced computational cost.
\end{abstract}

\pacs{05.10.Ln,87.15.H-,02.50.Ga} 

\maketitle

\section{Introduction}

Energy landscapes \cite{Reidys:02,Wales:03,Zhou:09} are a key concept for the
description of complex physical and biological systems. In particular, the
dynamics of structure formation (``folding'') of biopolymers, \eg{} protein or
ribonucleic acids, can be understood in terms of their energy landscapes
\cite{Anfinsen:73,Ming:08}. Formally, a landscape is determined by a set $X$
of micro-states (or conformations), a neighborhood structure of $X$ that encodes
which conformations can be reached from which other ones, and an energy function
$E:X\to\mathbb{R}$ which assigns an energy value to each state. In the case of
ribonucleic acids (RNA) it has been demonstrated that the dynamics of the
folding process can be captured in good approximation by merging large contiguous sets
of micro-states into macro-states
\cite{Flamm:02,Wolfinger:04a}. A typical mapping is in
terms of gradient basins: Each macro-state contains the micro-states from which
a given local minimum is reached by steepest descent in energy, including the
local minimum itself. The so-defined macro-states are also called inherent
structures in the  context of continuous disordered systems, see
ref.~\cite{Heuer:08} for a recent review.

Given a partitioning of the landscape, the dynamics is approximately described
as a Markov chain on the set of macro-states. In order to obtain this
description, the transition probabilities between macro-states in this Markov
chain need to be extracted from the original energy landscape.

As a first approximation, the {\em Arrhenius} equation predicts that the
transition probability is exponentially suppressed by the ratio between barrier
height and temperature. The barrier height (also called activation energy) from
minimum $a$ to minimum $b$ measures the minimal amount by which the system's
energy must increase along a path from $a$ to $b$
\cite{Flamm:02,Wolfinger:06a,Baiesi:09}. The accuracy of this approach is
limited because it ignores the multiplicity of low energy paths
\cite{Garstecki:99}. A more severe drawback is the complexity of computing
barrier height itself. For landscapes of RNA secondary structure
\cite{Fontana:93}, the problem is NP-hard \cite{Manuch:09,Thachuk:10}.

Commonly used methods
\cite{Flamm:02,Chan:94,Wuchty:99,Sibani:99,Burda:06}
for precise transition rate estimation are based on enumeration of all micro-states.
For landscapes of real combinatorial problems or long biopolymers with billions
of micro-states, however, enumeration is impractical with the given time
resources. Typically, limited storage capacity puts even more severe
restrictions on the size of tractable problems because a large fraction of the
enumerated micro-states needs to be kept in working memory. Some studies
partially circumvent this problem by considering only the low-energy fraction of
the landscape that is tractable with the available
resources~\cite{Sibani:93,Sibani:94,SchoenSibani:98,Wolfinger:06a}. Other
heuristic approaches \cite{Xayaphoummine:03,Tang:05,Thomas:05,Thomas:07,Geis:08a} restrict the
landscape to the subset of states likely to be traversed by certain
trajectories, \eg{} folding from the open chain to the ground state of a
biopolymer. 

Here we make a contribution to the original challenge of capturing an arbitrary
discrete landscape in terms of macro-states and transition probabilities. We
suggest a Markov chain Monte-Carlo sampling method for transition matrix
estimation. At difference with the earlier approaches, the memory requirement
scales linearly with the number of non-zero transition probabilities to be
determined. Other recent methods of stochastic landscape exploration
\cite{Gfeller:07,Prada:09} use trajectories of the original dynamics for
counting transitions between macro-states. In contrast, the idea behind the
present method is to explicitly explore boundaries between macro-states. To this
end, we confine the dynamics into a single macro-state $b$ and find and count
possible transitions from $b$ to all adjacent macro-states. This strategy allows
to select the regions of the landscape to be explored and the  precision to be
applied.

\section{Landscape and micro-state dynamics} \label{sec:landmic}

A discrete energy landscape is a triple $(X,E,M)$ where
\begin{itemize}
\item $X$ is a finite set of states,
\item $E:X \rightarrow {\mathbb R}$ is an energy function on $X$, and
\item $\ELneighX:X \rightarrow {\cal P}(X)$ is a neighborhood function or
``move set'' that assigns to each state $x \in X$ the set of its directly accessible
neighboring states. ${\cal P}(X)$ is the power set of $X$. Here
we assume that $\ELneighX$ is symmetric, \ie{} $x \in
\ELneigh{y} \Rightarrow y \in \ELneigh{x}$. By $\Delta$ we denote the
maximum number of neighbors, $\Delta = \max_{x \in X} |M(x)|$.
\end{itemize}

We consider a time-discrete stochastic dynamics on the state set $X$. Having
the Markov property, the dynamics is defined by giving the transition
probability $p_{x \rightarrow y}$ from each $x \in X$ to each
$y \in \ELneigh{x}$. Provided the system is in state
$x$ at time $t$, $p_{x \rightarrow y}$ is the probability that the system is in
state $y$ at time $t+1$. With probability $p_{x \rightarrow x} = 1 - \sum_{y \in
\ELneigh{x}} p_{x \rightarrow y}$, the system remains at state $x$.

Specifically, the Metropolis probabilities at inverse temperature $\beta$,
\begin{equation} 
p_{x \rightarrow y} = \Delta^{-1} \min\{ \exp(\beta[E(x)-E(y)]),1\} 
\end{equation} 
are used throughout this contribution. This
choice, however, is not compulsory. All that follows, and in particular the
estimation by sampling, applies to arbitrary choices of transition
probabilities leading to {\em ergodic} Markov chains. The ergodicity is
important because we need a unique stationary distribution $P(x)$ on $X$.

\section{Partitioning and macro-state dynamics} \label{sec:parti}

A partitioning of the landscape is a mapping $F$ from the set of micro-states
$X$ into a set of macro-states $B$.  Our goal here is to find a dynamics on $B$
that does have the Markov property while following the original micro-state
dynamics as closely as possible.  In general, however, a Markov chain is not
obtained as the direct mapping $(F(x_t))_{t=0}^\infty$ of a Markov chain
$(x_t)_{t=0}^\infty$ generated by the dynamics on $X$. The reason can be
sketched as follows. When the system is in a macro-state $b\in B$, the
probability of exiting to a macro-state $c$ depends on where exactly (in which
micro-state) the system is inside $b$. The micro-state assumed inside $b$,
however, depends on how the system entered $b$, which is again influenced by
the macro-state $a$ assumed before entering $b$.

Thus, the following simplifying assumption is made \cite{Kramers:40}. Given that the system is
found in macro-state $b \in B$, the micro-state $x \in X$ is distributed as
\begin{equation}\label{eq:pb}
P_b(x) = \left\{ \begin{array}{ll}
P(x) \; / \sum_{y \in F^{-1}(b)} P(y)~ & \textrm{if } x \in F^{-1}(b) \\
0                                      & \textrm{otherwise.}
\end{array} \right.
\end{equation}
This is the stationary distribution $P$ of the whole system restricted to
micro-states in $b$ and normalized appropriately.
Under this assumption, the probability of a transition to macro-state $c$,
when being in macro-state $b \neq c$ is
\begin{equation} \label{eq:qbc}
q_{b \rightarrow c} = \sum_{x \in F^{-1}(b)} \left( P_b(x)
\sum_{y \in \ELneigh{x} \cap F^{-1}(c)} p_{x \rightarrow y}\right)~.
\end{equation}
The inner sum is the probability of going to a micro-state $y$ belonging
to macro-state $c$ and being a neighbor of $x$, given that the system is
in state $x$. The outer sum represents the equilibrium weighting of 
the micro-states $x$ inside the given macro-state $b$. A straight-forward 
method determines the exact transition probabilities by performing the
sums in Eq.~(\ref{eq:qbc}), \ie{} exhaustive enumeration of all micro-states
and all neighbors \cite{Flamm:02,Sibani:99}.

Throughout this contribution, we consider the usual partitioning of $X$ with
respect to gradient basins but the method is not restricted to this choice. Two
micro-states $x,y \in X$ lie in the same macro-state $F(x)=F(y)$ if and only if
the steepest descent walks starting in $x$ and $y$ terminate in the same local
minimum. A state $u \in X$ is called local minimum, if $E(v)>E(u)$ for all $v
\in M(u)$. For a given landscape and partitioning, the macro-state transition
probabilities can be estimated by the sampling algorithm presented in the next
section.

\section{Sampling method}\label{sec:method}

The method we introduce computes an estimate of the transition probabilities $q$
in Eq.~(\ref{eq:qbc}) by a standard importance sampling restricted to a
macro-state $b$ using the micro-state probabilities $P_b(x)$ defined in
Eq.~(\ref{eq:pb}). Being in state $x_t \in F^{-1}(b)$ at time $t$, a
neighbor $z \in \ELneigh{x_t}$ is drawn at random with equal probabilities. The
suggestion is accepted as the next state, $x_{t+1}=z$, with probability
$\min\{1,P_b(z)/P_b(x_t)\}$. Otherwise the state remains the same,
$x_{t+1}=x_t$. This choice guarantees that the relative frequency of state $x$
tends towards the relative frequency $P_b(x)$
for increasing chain length $t \rightarrow \infty$ \cite{Metropolis:53}. 
For a realization of a Markov chain of
length $t_{\rm max}$, transition probabilities are estimated as
\begin{equation} \label{eq:qbcsampling}
q^\prime_{b \rightarrow c} = \frac{1}{t_{\rm max}} \sum_{t=1}^{t_{\rm max}}
\sum_{y \in \ELneigh{x_t} \cap F^{-1}(c)} p_{x_t \rightarrow y}~.
\end{equation}
In practice, the inner summation is performed only once at each time $t$,
because each neighbor $y$ of $x_t$ contributes to the transition probability to
exactly one  macro-state $F(y)$.

Computation time is saved by storing visited micro-states of basin $b$ and their
sets of neighbors with transition probabilities in a data structure with fast
search access, \eg{} in a hash table.  This is particularly advantageous in
cases with broadly distributed micro-state probabilities such as Boltzmann
weights at low temperature. Here the Markov chain will encounter the highly
probable (low energy) micro-states many times but neighbor sets and transition
probabilities are computed only once per state. In the usual cases where
macro-states are defined as basins of local minima, memory of visited states
also saves time in evaluating the macro-state assignment function $F$: When the
gradient walk starting at state $x$ reaches a micro-state known to be in basin
$b$, $x$ itself is known to belong to $b$. Thus in many cases the walk may be
terminated before reaching the ground state. Keeping previously visited
micro-states in memory, however, is not necessary for the method to work. It may
be handled according to the available resources. One may simply stop storing
micro-states when the designated memory has been filled.

So far we have described how to estimate probabilities of transitions from {\em
one} macro-state $b$ to others. The result is the $b$-th column vector
$(q^\prime_{b\rightarrow c})_{c\in B}$ of the
estimated transition matrix $q^\prime$ as given in Eq.~\ref{eq:qbcsampling}.
By applying the procedure separately to each
macro-state, the full matrix $q^\prime$ is obtained. This can be implemented
as an iterative exploration of the energy landscape without initial knowledge of
the set of macro-states. Whenever a neighbor $y$ of a state $x$ in the Markov
chain belongs to a macro-state $F(y)$ not previously seen, we add the pair
$(F(y),y)$ to a queue $Q$ of macro-states yet to work on. Initially, $Q$ may
contain only one particular pair $(b, x_0)$, \eg{} the completely unfolded state
$x_0$ of a polymer and the corresponding macro-state $b=F(x_0)$. The iterative
exploration of the landscape is implemented in the following loop. (i) Extract a
pair $(b, x_0)$ from $Q$; (ii) generate Markov chain inside $b$, starting at
$x_0$; (iii) obtain estimates according to Eq.~(\ref{eq:qbcsampling}) and add
newly discovered macro-states to $Q$; (iv) If $Q$ is not empty, resume at~(i).
Note, this method is directly parallelizable and will easily profit from
distributed computing. Several independent realizations of Markov chains with
respect to different macro-states can be run simultaneously, extracting from and
feeding to the same queue. An implementation of the method is part of
the Energy Landscape Library~\cite{Mann_ELL_BIRD07}.

\section{Number partitioning landscape} \label{sec:NPP}
The number partitioning problem (NPP) is a decision problem in the theory of
computation and computational complexity \cite{Garey:79,Mertens:01,Stadler:03}.
It asks if a given set $A$ of $N$ real non-negative numbers can be partitioned
into two subsets $B,C$ such that numbers in $B$ have the same sum as those in
$C$. In an equivalent formulation, we label the numbers in $A$ as
$a_1,\dots,a_N$ and use spin variables $x_1, \dots, x_N$ to encode if $a_i$ is
in subset $B$ ($x_i=+1$) or in subset $C$ ($x_i=-1$). This system has the set of
micro-states $X = \{-1,+1\}^N$. We define the energy of state $x\in X$ as
\begin{equation}
E(x) = | \sum_{i=1}^N x_i a_i | ~.
\end{equation}
Then the NPP amounts to the question if the ground state energy of this system
is zero.

The number partitioning {\em landscape} is obtained by using the hypercube as
the neighborhood structure. For each $x \in X$ we have
\begin{equation}
\ELneigh{x} = \{y \in X \;|\; d(x,y) =1 \}
\end{equation}
as the set of neighbors. The usual Hamming distance $d$ is used, so
$d(x,y)$ is the number of entries $i$ such that $x_i \neq y_i$. A local
move on the landscape means flipping one of the $N$ spin variables $x_i$.

Random instances are typically generated by drawing the $a_i$ as statistically
independent random variables uniformly distributed in the unit interval. Then
the expected number of local minima grows
exponentially with $N$, more precisely $\langle |B| \rangle \sim 2^N
N^{-3/2}$ \cite{Ferreira:98}. Here we use special instances of the NPP where
\begin{equation} \label{eq:nppspecial}
a_i = (i-1)^{-\alpha}
\end{equation}
with $\alpha =0.55$. For these instances, we have found the number of local
minima to grow exponentially with $N$ for $N \le 40$. However, the growth is
much slower than for randomly generated instances. At $N=40$, the
instance of Equation~(\ref{eq:nppspecial}) has $318$~local minima, to
be compared with an expected number of $\approx 10^{15}$~local minima for
randomly generated instances. Each landscape with $10\le N \le 40$
has at least one basin with an energy barrier $\ge 0.1$. We explore the
landscape at temperature $1/\beta=0.1$.

\begin{figure}
\begin{center}
\includegraphics[width=\columnwidth]{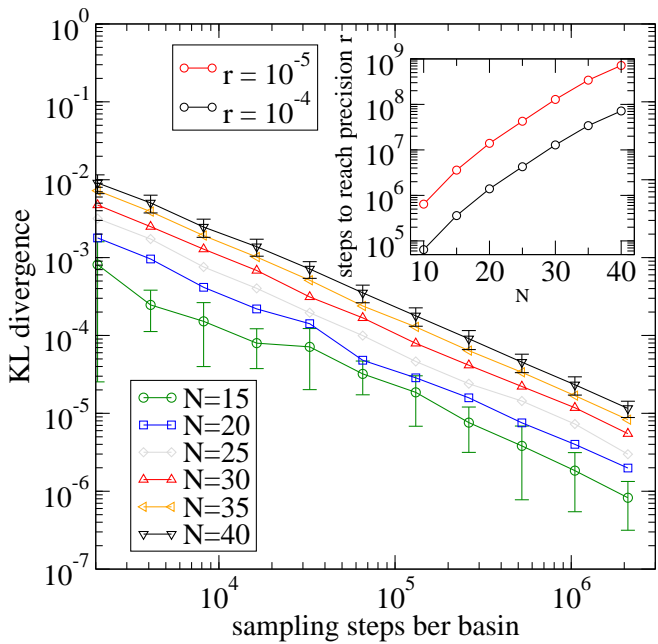}
\caption{\label{fig:kl_npp}
(Color online) The deviation of estimated transition probabilities from the exact values 
is inversely proportional to the number of sampling steps (main panel).
Shown are the analyzed special instances of number partitioning landscapes (see
Eq.~\ref{eq:nppspecial}) for various sizes~$N$. Error bars 
($N=15$ and $N=40$) indicate the standard deviation between errors for
different basins.  The inset shows the $N$-dependence of the total number of
sampling steps required for reaching a given precision, \ie{} lowering the error
below $r$. Given a macro-state $a$, we employ the Kullback-Leibler (KL)
divergence $D(.||..)$ \cite{Kullback:87} to define the error as
$\epsilon(s,a): = D(q^\prime(s,a) || q^\prime(2s,a))$, making a comparison
of the estimate of the outgoing transition probability vector 
$q^\prime(s,a) = (q^\prime_{a \rightarrow b})_{b \in B}(s)$ after $s$
sampling steps with its estimate after $2s$ sampling steps.
The plotted values are the equally weighted average of the errors
$\epsilon(s,a)$ over all macro-states $a \in B$.
}
\end{center}
\end{figure}

Figure \ref{fig:kl_npp} shows the convergence of the probability estimates. For
each system size $N$, the sampling error decreases inversely proportional to the
number of sampling steps performed per basin. Larger systems need more
computational effort to reach a certain precision. The inset of
Fig.~\ref{fig:kl_npp} indicates that the total computational effort required for
the error to fall below a given value grows sub-exponentially with $N$, to be
compared with a number of micro-states increasing as $2^N$. Thus under growing
$N$, sampling a strongly decreasing fraction of micro-states is sufficient in
order to reach a given precision.

\section{Folding landscape of an RNA switch} \label{sec:RNA}

\begin{figure}
\centerline{\includegraphics[width=\columnwidth]{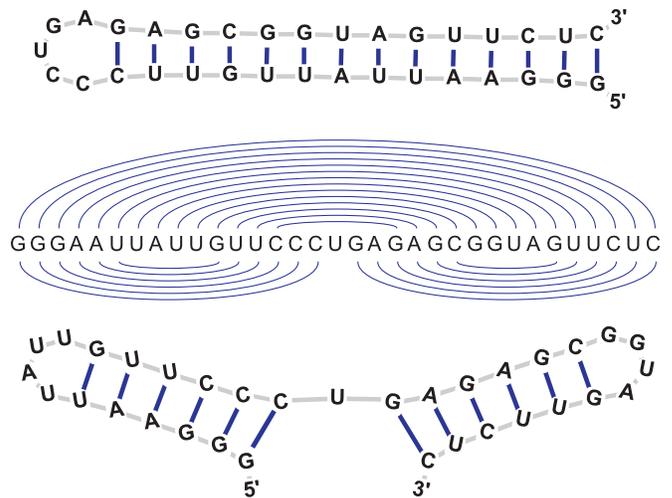}}
\caption{(Color online) RNA secondary structures (top/bottom) with energies
$-14.4${\tiny$\kcalmol$} and
$-14.3${\tiny$\kcalmol$} of the tested bistable RNA~d33 and their outer-planar linear Feynman diagrams
(middle) (drawn using jViz.Rna~v1.77~\cite{jViz-RNA:06}). Energy evaluation
and sequence design is based on Vienna RNA package v1.8.2~\cite{Hofacker:94a}
and the method from~\cite{Flamm:00b}.}
\label{fig:RNA_d33_graph}
\end{figure}

As a real-world example of folding landscapes of biopolymers we consider RNA
molecules~\cite{Flamm:00a}. The primary structure of an RNA molecule is a finite
sequence (a string) over the alphabet of the four nuclear bases $\{${\tt
A,C,G,U}$\}$. An RNA secondary structure is a list of pairs $(i,j)$ of positions
in the primary structure such that the following conditions hold. (1)~Base
combinations at pairing positions must be {\tt A-U} or {\tt G-C} (Watson-Crick
pairs) or {\tt G-U} (wobble pair); (2)~each position $i$ can pair with at most
one other position $j$; (3)~there are no two pairs $(i, j)$ and $(k, l)$ with $i
< k < j < l$. The latter condition forbids so-called pseudoknots and makes the
graph representation of a secondary structure outer-planar (see
Fig.~\ref{fig:RNA_d33_graph}).

In the folding landscape of an RNA sequence, the set of micro-states $X$
contains the valid secondary structures. The energy~$E(x)$ of a secondary
structure $x \in X$ is a sum over binding energies of stacks (contiguous regions
of binding) and entropic contributions from open (unbound) sections of the RNA
chain. For details of energy calculations, we refer to the
literature~\cite{Hofacker:94a,Tinoco:71,Freier:86}. Micro-states $x,y \in X$
are adjacent, \ie{} $y \in \ELneigh{x}$ and $x \in \ELneigh{y}$, if~$y$ can be
generated from~$x$ by adding or removing a single base pair. Shift
moves~\cite{Flamm:00a} are not considered in this contribution. When
the lowest energy neighbor of a structure is not unique the degeneracy
is resolved by the lexicographic ordering on string representations
of the structures
\cite{Flamm:02,Wolfinger:04a,Wolfinger:06a,Flamm:00a}.

Multistable RNAs, so called RNA-switches, are essential for the regulation of
cellular processes.  Thus, an understanding of the folding kinetics of such
molecules is of high importance. For a detailed overview see~\cite{Flamm:00b}.
Specifically, we work with the bistable RNA~d33 sequence shown in
Fig.~\ref{fig:RNA_d33_graph}. It has  $29,759,371$ micro-states, allowing for
full enumeration and thus for a comparative analysis with our method. Out of the
$3,223$ local minima, the two lowest are the secondary structures given in
Fig.~\ref{fig:RNA_d33_graph}. These two ground states have practically the same
energy. A walk between the ground states involves breaking all base pairs,
resulting in an energy barrier of height
$\Delta E = 1.18 \times 10^{-19}\text{ J}$.
The temperature for both sampling and energy calculation is
$T=(273.15+37.00)\text{ K}$. Therefore $1/\beta=k_B T = 4.28 \times
10^{-21}\text{ J}$ is more than one order of magnitude below the barrier height
$\Delta E$ of the RNA switch.

\begin{figure}
\begin{center}
\includegraphics[width=\columnwidth]{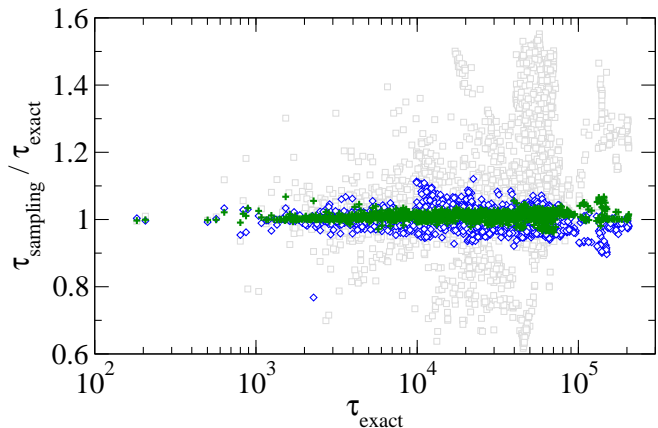}
\caption{\label{fig:tau_d33}
(Color online) Sampling precision in terms of the predicted average time $\tau(b)$ to reach
the ground state from a macro-state $b$ for RNA~d33. For each $b$, the
corresponding data point gives the ratio between $\tau_{\rm sampling}$ based on the sampled
transitions~$q'$ and the value $\tau_{\rm exact}$ from the exact ones~$q$ {\em
versus} $\tau_{\rm exact}$ itself. 
Symbols indicate number of sampling steps per macro-state as $10^3$ (squares),
$10^4$ (diamonds), and $10^5$ (crosses). The target set contains both
ground states.
}
\end{center}
\end{figure}

A comparison between exact and sampled transition probabilities is
made in terms of the average time $\tau(b)$ from macro-state $b$ to one of
the ground states. For a biopolymer as considered here, $\tau(b)$ is the
folding time when starting in an initial state $b$ such as the 
open chain. 

Given a set of target states $A \subset B$, the time to target is
$\tau(a)=0$ when starting in one of the target states $a \in A$ (boundary
condition). For a starting state
$b \in B \setminus A$, the average time $\tau(b)$ until first reaching one
of the target states obeys the recursion 
\begin{equation} \label{eq:taurecursion}
\tau(b) = 1+ \sum_{c \in B} q_{b \rightarrow c} \tau(c) ~.
\end{equation}
The average time to target from $b$ is one time step plus the time to target
from the state $c$ following $b$. The distribution of $c$ is given by the
transition probability $q_{b \rightarrow c}$. Time to target is also called exit
time \cite{Grimmett:82}.

Figure~\ref{fig:tau_d33} shows that $\tau(b)$-values based on
the sampled transition probabilities have small relative error for all
starting macro-states $b \in B$. With $10^4$ sampling
steps per basin, the ratios between exact and approximate times $\tau$
are in the range $[0.75;1.15]$. They fall into $[0.96; 1.07]$ when using
$10^5$ steps per basin.

To investigate the sampling error we compare, separately for each macro-state
$b$, the exact with the estimated transition probability vectors for leaving
$b$. We quantify the discrepancy between the two vectors by the Kullback-Leibler
divergence~\cite{Kullback:87}
\begin{equation} \label{eq:KL}
D(q^\prime_b || q_b) = \sum_{c \in B} q^\prime_{b \rightarrow c} \ln
\frac{q^\prime_{b \rightarrow c}}{q_{b\rightarrow c}}~.
\end{equation}

\begin{figure}
\begin{center}
\includegraphics[width=\columnwidth,clip]{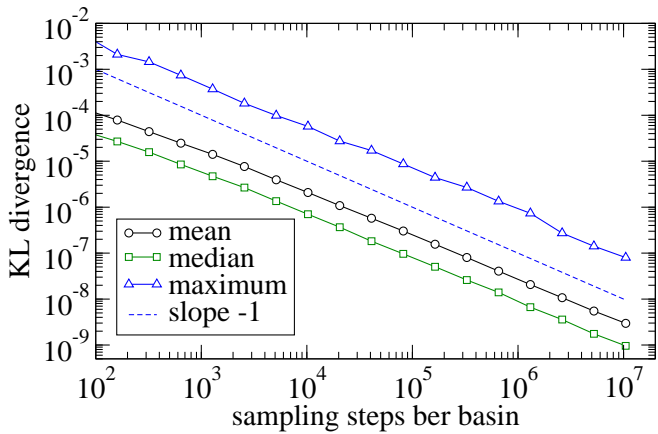}
\caption{\label{fig:kl_d33}
(Color online) Time evolution of the sampling error (KL divergence Eq.~\ref{eq:KL}) for the
folding landscape of the RNA switch molecule d33.
Mean, median, and maximum are for the
distribution of KL values over the $|B|=3223$ macro-states (local minima).
}
\end{center}
\end{figure}

Figure~\ref{fig:kl_d33} summarizes the evolution of the sampling error for
increasing sampling steps per basin. As in the number partitioning landscape,
Kullback-Leibler divergence (KL) decreases inversely proportional to the number
of sampling steps. The mean of the distribution of KL values across basins is
larger than the median by a factor of~3 indicating a broad distribution. This is
due to a broad distribution of macro-state sizes. Probability vectors for
macro-states comprising one or a few micro-states reach a low error after fewer
sampling steps than those for large macro-states. Still also the maximum of the
error across all basins decreases proportionally to the average. One of the
extensions of the method outlined in Sec.~\ref{sec:ext} chooses a number of
sampling steps individually for each macro-state based on its estimated
partition function.

\begin{figure}
\begin{center}
\includegraphics[width=\columnwidth]{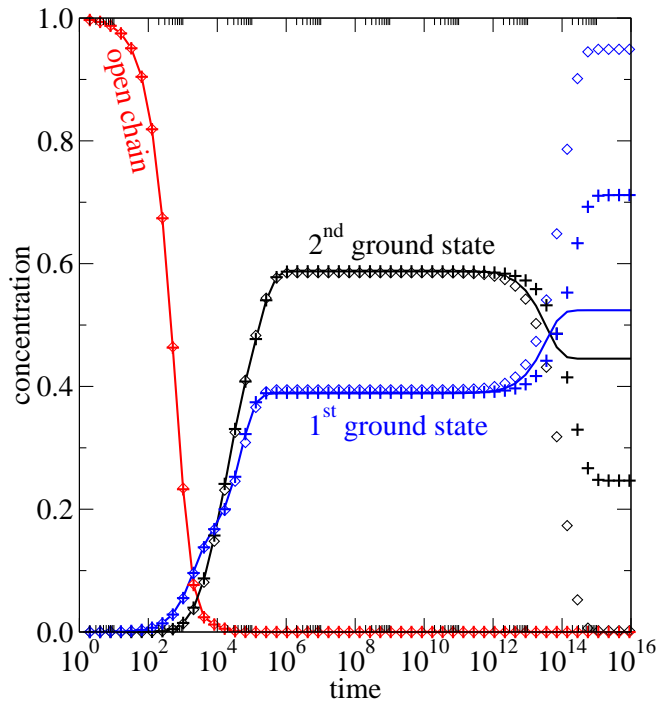}
\caption{\label{fig:dyn_d33}
(Color online) Time evolution of macro-state concentrations for RNA~d33 from the
exact transition probabilities~$q$ (curves) and from the estimates~$q^\prime$
via sampling for $10^4$ ($\diamond$) and $10^5$ steps per macro-state ($+$).
The concentrations both for the exact and the sampled transition rates are at
the stationary values for $t\ge 10^{15}$. See main text for discussion of
the discrepancy in these stationary concentrations.}
\end{center}
\end{figure}

In Figure~\ref{fig:dyn_d33}, we compare the kinetics of the molecule for the
approximated transition probabilities via sampling and the exact ones obtained
by enumeration of all micro-states of the landscape. As an initial condition we
choose the whole ensemble to be in the macro-state of the open chain (structure
without base pairs). As a qualitative description, the ensemble first populates
the first and, somewhat later, the second ground state. On an intermediate time
scale ($10^6$), an almost constant concentration vector is reached with the
second ground state dominating the first. However, this plateau concentration
vector is transient. Probability mass flows from the second to the first ground
state on a slow time scale ($10^{15}$) to reach the stationary
concentrations.

With transition probabilities obtained by sampling for $10^4$ steps per
macro-state, the kinetics is reproduced with high precision both in the timing
as well as the absolute concentration in the plateau where the relative error is
below $10^{-2}$. Since we hash the probabilities for already visited structures,
the computational effort per basin is dominated by the number of visited states
instead of overall sampling steps. Thus, small basins are sampled faster than
larger ones. When sampling $10^5$~steps per macro-state, which renders the
kinetics with even higher accuracy, computation time is still reduced by a
factor of $\approx 9$ compared with full enumeration.

The stationary concentrations found at time $t \ge 10^{15}$, however, do not
agree with the exact solution. A much larger number of sampling steps is
required to match these. Further tests with other RNA switch molecules yield the
same qualitative result for moderate number of sampling steps per basin. Both
the concentration levels and the time scales are faithfully reproduced by the
transition rates from sampling, except for the stationary concentrations.

A closer look at the particular structure of the landscape of RNA switches hints
at an explanation for the discrepancy. Both ground states have large and deep
basins. In RNA~d33, barriers to neighboring basins are all at least $8.6 \times
10^{-20}\text{ J}$ above ground state energy, which is more than $20\;k_B T$.
Exits from one ground state basin towards the other lead through a small number
of micro-states. When sampling the large ground state basins, these few salient
micro-states are likely to be missed. The subtle balance between incoming and
outgoing probability flow in the equilibrium is distorted. Due to the symmetry
of the move set, however, those missed non-zero transition probabilities can be
identified to some extent. When the forward rate is non-zero then the backward
rate must be non-zero as well. For dynamics with detailed balance, as considered
here, even quantitative correction of missed or undersampled rates is possible.
This is suggested as one of the extensions in the following section.

\section{Extensions and modifications of the method} \label{sec:ext}

We outline ideas for varying the method to potentially increase
efficiency and applicability in various settings. These
are not used in the applications in Sec.~\ref{sec:NPP} and~\ref{sec:RNA}.

\subsection{Guided sampling}
The stopping criterion (iv) of the outer loop for full transition matrix
estimation (Sec.~\ref{sec:method}) may be modified if we do not aim to explore
the whole landscape but only a subset of the set of macro-states
\cite{Burda:06}. Then $Q$ may be handled as a priority queue. For instance, we
may be interested only in transitions between macro-states below a certain
energy threshold or those involved in typical trajectories. In the latter case,
the next macro-state to be explored is the one that is reached from already
explored macro-states with the largest probability.

\subsection{Partition function estimation}

In addition to transition probabilities, the canonical partition function
\begin{equation} \label{eq:Zexact}
Z_b = \sum_{x \in F^{-1}(b)} \exp(-\beta E(x))
\end{equation}
of the macro-state~$b$ may be estimated at any time during the sampling.
Consider a subset $X \subseteq F^{-1}(b)$ of the micro-states of macro-state~$b$.
When sampling in the macro-state $b$, the fraction of time the Markov chain
spends in $X$ is
\begin{equation} \label{eq:timefrac}
r = \frac{1}{Z_b} \sum_{x \in X} \exp(-\beta E(x))~.
\end{equation}
Therefore $Z_b$ can be calculated when knowing $r$ and the energies of all
states in $X$. Now $X$ can be taken as the set of states visited in the first
$t^\ast$ steps of the Markov chain, $X = \{ x_t \;|\; t=1,\dots,t^\ast \}$.
An estimate $r^\prime$ of $r$ is obtained by counting how often the Markov
chain visits states in $X$ during a sufficiently long time interval
$[t_\text{start}, t_\text{stop}[$,
\begin{equation}
r^\prime = \frac{| \{\; t \;|\; t_\text{start} \le t < t_\text{stop}
\;\wedge\; x_t \in X \} | } {t_\text{stop} - t_\text{start}}~.
\end{equation}
In order to obtain an unbiased estimate of $r$, this time interval must
not overlap with the time steps during which $X$ is recorded, thus
$t_\text{start} > t^\ast$. By solving Eq.~(\ref{eq:timefrac}) for $Z_b$
and replacing $r$ with the estimate $r^\prime$, we obtain
\begin{equation}
Z_b^\prime = \frac{1}{r} \sum_{x \in X} \exp(-\beta E(x))
\end{equation}
as an unbiased estimate of the partition function $Z_b$.

\subsection{Sampling time adjustment}

The estimate $r(t)$ may also be used for adapting the length of the Markov chain
exploring macro-state $b$ to the size of $b$. The sampling will be run until the
fraction of covered probability mass exceeds a certain threshold, \eg{} stopping
as soon as $r(t)>0.5$.

\subsection{Landscape coarse-graining}

The state space can be coarse-grained beyond the initially chosen macro-state
state partitioning by dynamically merging macro-states \cite{Bongini:09}. For
merging macro-state $b$ into macro-state $a$, the affected entries in the matrix
$q$ are replaced by
\begin{eqnarray}
\hat{q}_{a \rightarrow c} & = &
\frac{ Z_a q_{a \rightarrow c} + Z_b q_{b \rightarrow c} }{Z_a+Z_b}\\
\hat{q}_{c \rightarrow a} & = & q_{c \rightarrow  a} + q_{c \rightarrow b}
\end{eqnarray}
for all macro-states $c \notin \{ a,b \}$. The new diagonal element
$\hat{q}_{a \rightarrow a}$ is obtained by normalization of probability.
The row and column of macro-state $b$ are set zero (or deleted). In a
separate index, the mapping of macro-state $b$ to macro-state $a$ is stored.

Strategies for the choice of macro-states to be merged need to be explored
yet. A reasonable starting point is to choose pairs of macro-states with
high overlap in successor states, \eg{} choosing $a$ and $b$ such that
\begin{equation}
\sum_{c \in B} q_{a \rightarrow c} q_{b \rightarrow c} 
\end{equation}
is maximal. 

\subsection{Balancing the transition probability matrix}
If the micro-state dynamics in terms of the transition probabilities 
$p_{x \rightarrow y}$ fulfills detailed balance, then so does the
macro-state dynamics with transition probabilities $q_{b \rightarrow c}$.
Detailed balance means
\begin{equation} \label{eq:detailed_balance_Z}
Z_l \; q_{l\rightarrow s} = Z_s \; q_{s \rightarrow l}~.
\end{equation}
for all macro-state pairs $(l,s)$. The transition probabilities
$q^\prime_{l\rightarrow s}$ obtained by the sampling, however, need not
fulfill the same condition. By the transformation
\begin{equation} \label{eq:balancing}
q^\ast_{l \rightarrow s} = \frac{1}{2} q^\prime_{l \rightarrow s}
+ \frac{1}{2} q^\prime_{s \rightarrow l} \frac{Z_s}{Z_l}
\end{equation}
transition probabilities $q^\ast$ with detailed balance are obtained.
The transformation also serves to impose a known stationary distribution
of concentrations on the transition probability matrix.\\

\section{Conclusion and discussion} \label{sec:concl}
 
When coarse-graining the state space of an energy landscape into macro-states,
transition probabilities between macro-states have to be obtained in order to
capture the coarse-grained stochastic dynamics. Here we have introduced a
sampling method that allows for a fast yet accurate estimation of these
transition probabilities. We have demonstrated the scalability of the approach with system
size for special instances of the number partitioning problem. As a real-world
application, we have analyzed the folding landscape of the secondary structure
of an RNA switch as an example of a biopolymer. Its rich dynamic behavior on
separate fast and slow timescales is accurately rendered by transition
probabilities obtained with low computational cost.

The general method introduced here may serve as a flexible framework for
stochastic exploration of energy landscapes. As laid out in the
Sec.~\ref{sec:ext}, several extensions and modifications may be made to obtain
increased performance and wider applicability. In particular, the high variation
of macro-state sizes may be exploited in a scheme for an automatic choice of
sampling effort. Furthermore, the merging of small macro-states with larger
neighbors during the sampling may lead to more manageable and potentially more
meaningful partitions of the landscape akin to metabasins~\cite{Heuer:08}.

In ongoing and future work, the method shall be applied to other energy
landscapes including those of state-discrete protein folding
dynamics~\cite{Dill:85,Mann_2008,Mann_2008_2}. Such landscapes have been shown
to be amenable to sampling approaches~\cite{Wust:08}. Another field of
application of our method is the clarification of concepts for dynamics on
energy surfaces, such as the notion of a folding
funnel~\cite{Leopold:92,Garstecki:99,Klemm:08}.

\acknowledgments
KK gratefully acknowledges funding from VolkswagenStiftung.


\bibliography{mann_klemm_sampling}

\end{document}